\documentclass{vgtc}                    

\usepackage{mathptmx}
\usepackage{times}
\usepackage{epsfig}
\usepackage{graphicx}
\usepackage{amsmath}
\usepackage{amssymb}
\usepackage[linesnumbered,ruled,vlined]{algorithm2e}
\usepackage{booktabs}
\usepackage{array}

\onlineid{0}

\vgtccategory{Research}

\vgtcinsertpkg

\title{WebVRGIS Based City Bigdata 3D Visualization and Analysis}

\author{Xiaoming Li$^{a,b,c}$, Zhihan Lv$^{a}$, Baoyun Zhang$^{a}$, Weixi Wang$^{b,c}$, Shengzhong Feng$^{a}$, Jinxing Hu$^{a}$\\
a. Shenzhen Institutes of Advanced Technology(SIAT), Chinese Academy of Science, China\\
b. Shenzhen Research Center of Digital City Engineering, Shenzhen, China\\
c. Key Laboratory of Urban Land Resources Monitoring and Simulation, \\Ministry of Land and Resources, Shenzhen, China\\
lvzhihan@gmail.com, xm.li@siat.ac.cn, jx.hu@siat.ac.cn, sz.feng@siat.ac.cn} 

\abstract{This paper shows the WEBVRGIS platform overlying multiple types of data
   about Shenzhen over a 3d globe. The amount of information that can be
   visualized with this platform is overwhelming, and the GIS-based
   navigational scheme allows to have great flexibility to access the
   different available data sources. For example, visualising
   historical and forecasted passenger volume at stations could be very
   helpful when overlaid with other social data.} 

\CCScatlist{ 
 \CCScat{1.3.7}{Computer Graphics}{Three-Dimensional Graphics and Realism}{Virtual reality}
}

\begin{document}

\maketitle

\begin{figure}
    \begin{center}
    \includegraphics[width=0.9\columnwidth]{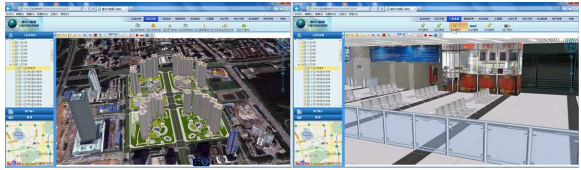}
    \caption{The UI of the proposed system.}
    \label{fig:3dui6}
        \end{center}
\end{figure}

\firstsection{Introduction}


Nowadays, there is an increasing interest in creating Virtual Reality Geographical Information System (VRGIS), which can obtain the landscape
geospatial data dynamically and perform rich visual 3D analysis, calculations, managements based on Geographical Information System (GIS) data. The web version of VRGIS is so-called WebVRGIS. A parallel trend, the utilize of bigdata is becoming a hot research topic rapidly recently. GIS data
has several characteristics, such as large scale, diverse
predictable and real-time, which falls in the range of definition
of Big Data~\cite{bigdata}. Besides, to improve the accuracy of modeling, the city planning has an increasingly high demand for
the realistic display of VR system, however this will inevitably
lead to the growth of the volume of data. Virtual scene from a single building to the city scale is also resulting
in the increased amount of data. In addition, beside spatial data integration, new user interfaces for geo-databases is also expected~\cite{Breunig:2011:RGR:1998664.1998871}.
Therefore, the management and development of city big data using virtual reality technology is a promising and inspiring approach. 
As a practical tool, most commonly used functions of VRGIS are improved according to practical needs~\cite{Lin201374}. In our platform, all the presented functions are extracted from the practical customer needs. Some early related systems from both academy and industry have inspired our work~\cite{Zhang:2007:DDD:1279008.1279148}~\cite{Ma:2010:IVN:1869057.1869058}.

\begin{figure}
    \begin{center}
    \includegraphics[width=0.9\columnwidth]{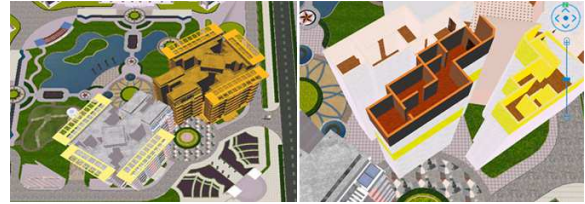}
    \caption{Left: 3D Planning model generation; Right: Selected model}
    \label{fig:3dui1}
        \end{center}
\end{figure}
\begin{figure}
    \begin{center}
    \includegraphics[width=0.9\columnwidth]{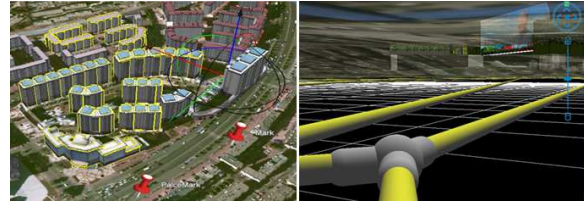}
    \caption{Left: Above-ground city; Right: Underground pipelines data}
    \label{fig:3dui2}
        \end{center}
\end{figure}

\section{System}

In this paper, we utilize 3D Shenzhen as a convincing case to present WebVRGIS~\cite{lviconip}, which is based on WebVR engine~\cite{lv2011webvr}. Shenzhen is a new city, however, it has the highest population density in China. It causes some embarrassments to the city information management.
As shown in figure~\ref{fig:3dui6}, the presented platform can assistant the social service agencies to make full use of Virtual Reality and database technology to improve the comprehensive level of the city's service management. To achieve shares of information resources of all departments and the dynamic tracking for the population and companies, geospatial information by constructing an integrated information platform of social services which are the area of coverage, street, four levels of administrative network.

The 3D visualization of city building is to take inventory and display various types of object data management and resource data within the community area which is so-called virtual community~\cite{lu2013webvrgis}, and thus help the social service agencies to grasp the work base by the supporting of holographic foundation library, such as population and companies, houses, events and urban component and other factors. Based on RIA technology~\cite{zhang2009research}, it can achieve holographic data query for every household of the specific community respectively by building 3D virtual house and associate it with relevant data.

\begin{figure}
    \begin{center}
    \includegraphics[width=1\columnwidth]{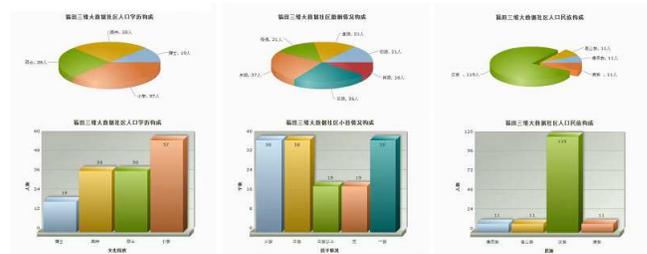}
    \caption{Education background composition; Children age composition; Nationality composition}
    \label{fig:3dui5}
        \end{center}
\end{figure}

The geographic statistical analysis is to assist management decision-making and conduct data analysis. It can provide decision support for social service agencies by the supporting of management, services, themes library of the work and the library of task extension, as well as holographic foundation library，such as population and companies, houses, events, urban component. It can construct various application scenarios query mode based on the jurisdiction area, and community or the range of grid query statistics. To provide a presentation way of table and graphics，such as charts of pie, line, area，scatter, bubble, radar, histogram and normal distribution geographic bitmap(map dotted) or thermodynamic, as shown in figure~\ref{fig:3dui5}.

\begin{figure}
    \begin{center}
    \includegraphics[width=0.9\columnwidth]{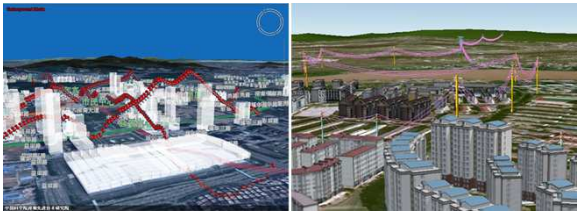}
    \caption{Left: Subway route tracking; Right: City power grid}
    \label{fig:3dui3}
        \end{center}
\end{figure}

\section{Graphic User Interface}

With a 3D earth model as the browser, this system is loaded with all 3D model data and the 3D visualization analysis result. By zooming in the camera horizon, it is possible to see the detailed level of buildings. By selecting the house inquiry, it is possible to find out the information of the address and owner, and to locate the house in the 3D scene. 3D roaming function can not only conduct soaring top view observation above the community, but can also observe the detailed layout near the street, and further enter the building to observe the internal building structure. In addition to the observation of above-ground model, it can also observe and manage the underground data under the city ground, as shown in figure~\ref{fig:3dui2}.

On the website part, the top view-based position observation is conducted through a 2D hawk eye map on the left. The tool bar on the top of the earth browser contains 3D GIS analysis function and urban data management, analysis and visualization. The tree-view management control on the upper left contains various layers in the earth browser. By clicking the tree view management control on the left, it is possible to display the 3D model of each building and each room respectively. 

Community analysis function can have a comprehensive observation to the 3D model of a community and analyze and display various data, including population age composition, education background composition, nationality composition, community marriage status and employment status. This platform can manage the urban traffic and real-time road condition. It can visualize the road condition in the form of line and plane. In addition to real-time data, it can also load the historical data. 

\section{Implementation}

The 3D rendering layer of the platform is based on our WebVRGIS engine which is developed by C++, OpenGL and wrapped into ActiveX plugin~\cite{lviconip}. The city bigdata contains the land and ocean~\cite{su20143d}~\cite{lv20143d}, the above-ground and underground, outdoor and indoor, building and people, real-time and history as well as the forecast. In order to solve the issue of massive 3D geo-database dynamic updating, an effective Spatiotemporal database model (SDM) based on event, semantics, and state is employed in the system. Spatiotemporal visualization has a natural advantage 
of revealing overall tendencies and movement patterns~\cite{zhong2012spatiotemporal}.

\section{Conclusion}

The 3D Shenzhen case proves 3D city visualization and analysis platform is a useful tool for the social service agencies and citizens for browsing
and analyzing city big data directly, and is agreed upon as being immediately useful. Some novel interaction approaches are considered to integrate in our future work~\cite{lv2013wearable}.

\section*{Acknowledgments}
The authors are thankful to the National Natural Science Fund for the Youth of China (41301439), LIESMARS Open Found(11I01) and Shenzhen Scientific \& Research Development Fund (JC201105190951A, JC201005270331A, CXZZ20130321092415392).

\bibliographystyle{abbrv}
\bibliography{VRVIDEO}

\begin{thebibliography}{10}

\bibitem{Breunig:2011:RGR:1998664.1998871}
M.~Breunig and S.~Zlatanova.
\newblock Review: 3d geo-database research: Retrospective and future
  directions.
\newblock {\em Comput. Geosci.}, 37(7):791--803, July 2011.

\bibitem{bigdata}
F.~Briggs.
\newblock Large data - great opportunities.
\newblock Presented at IDF2012, Beijing, 2012.

\bibitem{Lin201374}
H.~Lin, M.~Chen, G.~Lu, Q.~Zhu, J.~Gong, X.~You, Y.~Wen, B.~Xu, and M.~Hu.
\newblock Virtual geographic environments (vges): A new generation of
  geographic analysis tool.
\newblock {\em Earth-Science Reviews}, 126(0):74 -- 84, 2013.

\bibitem{lu2013webvrgis}
Z.~Lu, S.~U. Rehman, and G.~Chen.
\newblock Webvrgis: Webgis based interactive online 3d virtual community.
\newblock In {\em Virtual Reality and Visualization (ICVRV), 2013 International
  Conference on}, pages 94--99. IEEE, 2013.

\bibitem{lv2013wearable}
Z.~Lv.
\newblock Wearable smartphone: Wearable hybrid framework for hand and foot
  gesture interaction on smartphone.
\newblock In {\em 2013 IEEE International Conference on Computer Vision
  Workshops}, pages 436--443. IEEE, 2013.

\bibitem{lviconip}
Z.~Lv, S.~Réhman, and G.~Chen.
\newblock Webvrgis: A p2p network engine for vr data and gis analysis.
\newblock In M.~Lee, A.~Hirose, Z.-G. Hou, and R.~Kil, editors, {\em Neural
  Information Processing}, volume 8226 of {\em Lecture Notes in Computer
  Science}, pages 503--510. Springer Berlin Heidelberg, 2013.

\bibitem{lv20143d}
Z.~Lv and T.~Su.
\newblock 3d seabed modeling and visualization on ubiquitous context.
\newblock In {\em SIGGRAPH Asia 2014 Posters}, page~33. ACM, 2014.

\bibitem{lv2011webvr}
Z.~Lv, T.~Yin, Y.~Han, Y.~Chen, and G.~Chen.
\newblock Webvr--web virtual reality engine based on p2p network.
\newblock {\em Journal of Networks}, 6(7), 2011.

\bibitem{Ma:2010:IVN:1869057.1869058}
C.~Ma, G.~Chen, Y.~Han, Y.~Qi, and Y.~Chen.
\newblock An integrated vr\&ndash;gis navigation platform for city-region
  simulation.
\newblock {\em Comput. Animat. Virtual Worlds}, 21(5):499--507, Sept. 2010.

\bibitem{su20143d}
T.~Su, Z.~Lv, S.~Gao, X.~Li, and H.~Lv.
\newblock 3d seabed: 3d modeling and visualization platform for the seabed.
\newblock In {\em Multimedia and Expo Workshops (ICMEW), 2014 IEEE
  International Conference on}, pages 1--6. IEEE, 2014.

\bibitem{Zhang:2007:DDD:1279008.1279148}
J.~Zhang, J.~Gong, H.~Lin, G.~Wang, J.~Huang, J.~Zhu, B.~Xu, and J.~Teng.
\newblock Design and development of distributed virtual geographic environment
  system based on web services.
\newblock {\em Inf. Sci.}, 177(19):3968--3980, Oct. 2007.

\bibitem{zhang2009research}
M.~Zhang, Z.~Lv, X.~Zhang, G.~Chen, and K.~Zhang.
\newblock Research and application of the 3d virtual community based on webvr
  and ria.
\newblock {\em Computer and Information Science}, 2(1):P84, 2009.

\bibitem{zhong2012spatiotemporal}
C.~Zhong, T.~Wang, W.~Zeng, and S.~M. Arisona.
\newblock Spatiotemporal visualisation: A survey and outlook.
\newblock In {\em Digital Urban Modeling and Simulation}, pages 299--317.
  Springer, 2012.

\end{thebibliography}
\end{document}